\documentclass[aps,prb,reprint,amsmath,amssymb,groupedaddress]{revtex4-2}

\usepackage{graphicx}
\usepackage{dcolumn}
\usepackage{bm}
\usepackage{color}
\usepackage{siunitx}
\usepackage{todonotes}
\usepackage{threeparttable}
\usepackage{gensymb}

\definecolor{imcolor}{rgb}{0.5,0.,0.5}				

\definecolor{ohcolor}{rgb}{0.,0.5,0.5}

\definecolor{hpcolor}{rgb}{0.5,0.5,0}

\begin{document}

\title{First-principles study of magnetic states and  the anomalous Hall conductivity of  $M$Nb$_3$S$_6$ ($M$=Co, Fe, Mn, and Ni)}

\author{Hyowon Park$^{1,2}$, Olle Heinonen$^{1}$, and Ivar Martin$^{1}$}

\affiliation{$^1$Materials Science Division, Argonne National Laboratory, Argonne, IL, 60439, USA,\\
$^2$Department of Physics, University of Illinois at Chicago, Chicago, IL 60607, USA }

\date{\today}

\begin{abstract}
Inspired by the observation of the extremely large anomalous Hall effect in the absence of applied magnetic fields or uniform magnetization in CoNb$_3$S$_6$ [Nature Comm. {\bf9}, 3280  (2018); Phys.  Rev.  Research {\bf2}, 023051 (2020)], 
we perform a first-principles study of this and related compounds of the $M$Nb$_3$S$_6$ type with different transition metal $M$ ions to determine their  magnetic orders and the anomalous Hall conductivity (AHC). We find that non-coplanar antiferromagnetic  ordering is favored relative to collinear or coplanar order in the case of $M$=Co, Fe and Ni, while ferromagnetic ordering is favored in MnNb$_3$S$_6$ at low temperatures. The AHC in these materials with non-coplanar spin ordering can reach about $e^2/h$ per crystalline layer, while being negligible for coplanar and collinear cases.
We also find that the AHC depends sensitively on doping and reaches a maximum for  intermediate values of the local spin exchange potential between 0.3 and 0.8 eV. Our AHC results are  consistent with the reported Hall measurements in CoNb$_3$S$_6$ and suggest a possibility of similarly large anomalous Hall effects in related compounds.
\end{abstract}

\maketitle

\section{Introduction}

Despite having a simple semiclassical explanation, the Hall effect \cite{hall_new_1879} is a remarkable quantum phenomenon that is caused by  a subtle influence of the electromagnetic vector potential  on the quantum mechanical phase of electrons. In weak magnetic fields, the transverse Hall voltage, being determined by the Lorenz force, is proportional both to the electric current and the magnetic field. However, in the limit of strong magnetic fields, deviations from this simple dependence become  prominent, culminating  in two-dimensional systems as quantized values of Hall conductance \cite{klitzing_new_1980}, precise enough to serve as a universal standard \cite{Jeckelmann_2001}.

Reaching quantized levels of the Hall conductance in the original experiments \cite{klitzing_new_1980,paalanen1982quantized} required a combination of low carrier density and high magnetic fields. The reason is that the quantum Hall effect occurs when the electronic density $n_{2d}$ and quantum flux density, $n_\phi = B/\phi_0 = 2eB/h$ are comparable. 
Thus, reaching the quantized limit in conventional materials with their high carrier density would require extremely high magnetic fields, $10^3$~T and above.
This might seem to imply that bulk materials are destined to remain in the semiclassical -- weak-field -- limit of the Hall effect. This is however not true, and the reason is that electrons in solids experience additional interactions that can have a dramatic effect on their quantum mechanical phase. These are exchange interactions with magnetic moments of atoms and spin-orbit interactions. Both  can  influence the Bloch wave functions of electrons, resulting in nontrivial features in bands structures, leading to nonzero Hall effect even in the absence of external  magnetic field  \cite{chang1995berry,chang1996berry,sundaram1999wave,nagaosa_anomalous_2010, xiao_berry_2010}. A classic example is the anomalous Hall effect (AHE) in ferromagnetic (FM) iron \cite{kondo1962anomalous,yao2004}, where magnetic ordering with the help of spin-orbit interaction can deflect moving electrons in a way similar to the conventional \textit{orbital} magnetic field. A more exotic possibility, which does not require explicit spin-orbit coupling, relies on the interaction between spins of itinerant electrons and magnetic moments of atoms ordered into non-coplanar states \cite{taguchi_spin_2001, shindou_orbital_2001, martin_scalar_2008, machida_time-reversal_2010}. In this case, the Berry phase imprinted on electrons can reach values that are equivalent to having a flux quantum piercing the magnetic unit cell, which translates into an extremely large effective magnetic field. Notably, the magnetic ordering itself can be purely antiferromagnetic (AFM), that is, lacking any net magnetic moment.  Moreover, the AHE may persist even if there is no static magnetic order, but only finite scalar chirality \cite{machida_time-reversal_2010, kato_stability_2010}.

In recent years, several materials have been experimentally discovered that show complex AFM ordering and a very large AHC in the absence of magnetic fields or uniform magnetization \cite{nakatsuji_large_2015-1, nayak_large_nodate, zhang_strong_2017, ikhlas_large_2017, ghimire_large_2018, kimata_magnetic_2019, ghimire_competing_2020, asaba_anomalous_2020}. Despite these successes, observing a {\em quantized} AHC in these systems has remained elusive. 
Even if a system has topologically non-trivial bands that could lead to quantization \cite{thouless_quantized_1982}, accidental doping or magnetic domain structure may preclude its experimental observation \cite{tenasini_giant_2020}.

\begin{figure}[!ht]
\includegraphics[width=1.02\linewidth]{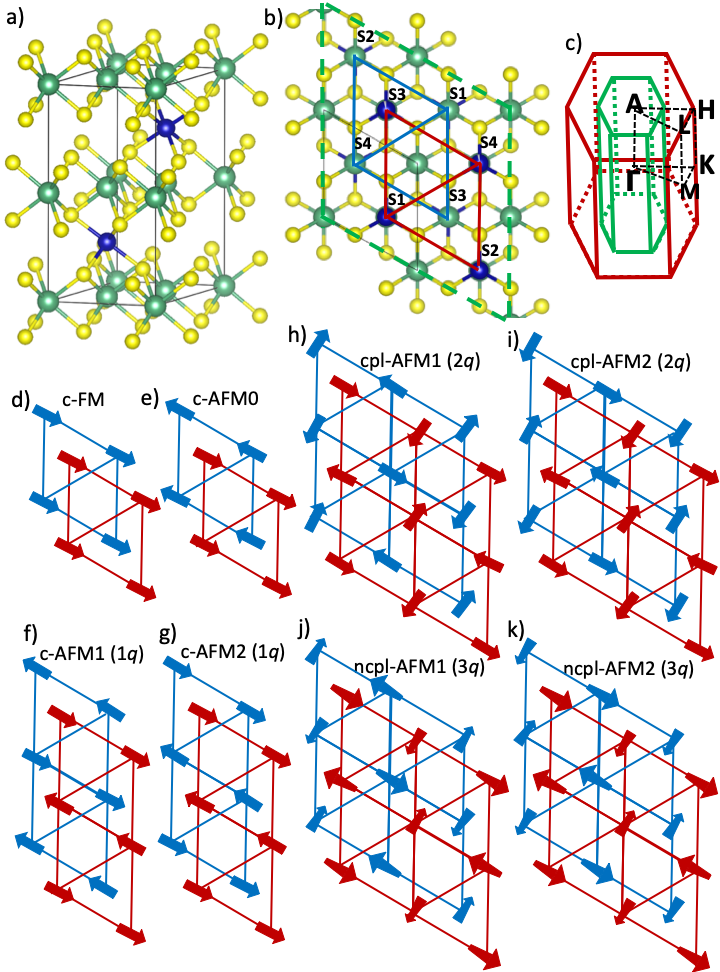}
\caption{a) The side view of the crystal structure of $M$Nb$_3$S$_6$ with $M$=Mn, Fe, Co, and Ni plotted using the VESTA package~\cite{vesta} ($M$: blue, Nb: green, S: yellow). b) The top view of the crystal structure. Solid line: primitive unit cell. Dashed line: magnetic unit cell for non-collinear states. Four different spins are labeled as S1, S2, S3, and S4 in each layer (top layer: red, bottom layer: blue). c)  The Brillouin zone of the primitive unit cell (red) and the magnetic unit cell (green). The high-symmetry points are used for the $k-$path of band structure plots. d-g) The collinear FM/AFM spin structures. h-k) The non-collinear (coplanar (cpl-) and non-coplanar(ncpl-)) AFM spin structures. In the ncpl-AFM case, the large (small) arrow head means the spin is pointing upward (downward) along the $c$-axis.
}
\label{fig:struct}
\end{figure}

A system that is perhaps one of the most promising candidates to exhibit a quantized AHC is CoNb$_3$S$_6$.  It has the layered hexagonal structure of NbS$_2$, intercalated by triangular layers of Co \cite{anzenhofer}. Based on early neutron scattering data, the magnetic structure was believed to be a simple collinear AFM \cite{parkin_magnetic_1983} (see Fig.~\ref{fig:struct}f,g). So it came as a big surprise when an extremely large AHE,  with the Hall conductivity comparable to $e^2/h$ per structural layer, was recently discovered in this system \cite{ghimire_large_2018, tenasini_giant_2020}. Indeed, collinear AFMs are invariant under a combination of lattice translation and time-reversal; however, since time-reversal flips the sign of Hall effect and translation leaves it unchanged, the AHE must vanish and, therefore, such AFM order cannot be responsible for the observed anomalous Hall response. More recently, neutron scattering experiments were performed \cite{zaharko} on the new samples of CoNb$_3$S$_6$ and the results did not exclude more complex magnetic orderings, in addition to the collinear AFM.  
The early neutron scattering measurement by Parkin $et$ $al$~\cite{parkin_magnetic_1983} reported the primary peak positions of CoNb$_3$S$_6$ at the $M$ points of the Brillrouin zone and their results can be interpreted as either a multi-domain $1q$ spin ordering with all three possible directions of $q$ present, or a single-domain $3q$ spin ordering. Distinguishing between these two possibilities require detailed analysis of higher-order scattering peaks that had not been done at the time. Although it is difficult to identify the magnetic order in these compounds experimentally and noting that a non-zero AHE could be possible even without any static magnetic orders, we focus here on the possibility of the AHE in CoNb$_3$S$_6$ and related compounds originating from static non-coplanar magnetic orders. Such orders can be studied using first-principles calculations.
One possibility that was not ruled out by the experiment is a ``tetrahedral" noncoplanar AFM  state. In a given triangular Co plane, the ordering simultaneously involves three symmetry-related AFM wavevectors, in contrast to the simple collinear AFM state, which only has one. The tetrahedral order has four spins per magnetic unit cell, pointing towards corners of a regular tetrahedron (see Fig.~\ref{fig:struct}j-k).
This state was theoretically proposed as a route to quantized AHC \cite{shindou_orbital_2001, martin_scalar_2008}---even in the absence of band spin-orbit interactions.  

In order to provide first-principles-based insights and to guide 
future experiments, here we computationally explore possible magnetic states, band structure, and AHC of CoNb$_3$S$_6$ and related materials where Co is replaced by other transition metal atoms, Fe, Ni, and Mn. 
Remarkably, we find that the non-coplanar AFM order  has the lowest energy in all cases except Mn. 
The AHC result calculated from the magnetic band structure of the non-coplanar AFM ordering shows that a sizable Hall conductivity can be obtained in $M$Nb$_3$S$_6$ for the $M=$ Co and Fe cases. Within out calculations, we find that all materials are gapless, which allows AHC to vary as a function of  doping and local spin exchange interaction. 

Our paper is organized as follows. In Sec.$\:$\ref{sec:method}, we describe the computational methods that are used throughout the paper. In Sec.$\:$\ref{sec:magnetism}, we compare the energies of various collinear and non-collinear magnetic states of $M$Nb$_3$S$_6$. In Sec.$\:$\ref{sec:bands}, we construct and compare the  non-magnetic, collinear, and non-collinear band structures   and in Sec.$\:$\ref{sec:AHC} we present  the AHC results for the non-coplanar spin states  (obtained at different dopings and local exchange  interactions). We conclude with a discussion in Sec.$\:$\ref{sec:conclusion}.

\section{Computational Methods}
\label{sec:method}

We adopt the Vienna Ab-initio Simulation Package (VASP)~\cite{vasp1,vasp2} code for the density functional theory (DFT) calculations of $M$Nb$_3$S$_6$ compounds.
The crystal structure of $M$Nb$_3$S$_6$ compounds has the group symmetry under the space group $P6_322$ and the lattice parameters and atomic positions have been measured using the X-ray diffraction experiment for $M$= Mn, Co, Fe and Ni~\cite{anzenhofer}.
We use the experimental structural parameters of $M$Nb$_3$S$_6$ for the DFT calculations and 
pseudo-potentials generated from the projector augmented wave method~\cite{PAW} with the valence electron configurations of $3d^64s^1$ (Mn), $3d^74s^1$ (Fe), $3d^84s^1$ (Co), $3d^94s^1$ (Ni), $4p^64d^45s^1$ (Nb), and $3s^23p^4$ (S).
The Perdew-Burke-Ernzerhof (PBE)~\cite{PBE_functional} functional is used for the exchange-correlation functional.
We use the energy cut-off for the plane-wave basis as 400~eV and a $8\times8\times4$ $k-$point grid for the primitive cell of $M$Nb$_3$S$_6$.
To compute the magnetic states of collinear and non-collinear AFM, we adopt the $2\times2\times1$ magnetic cell extended from the primitive cell. 
The same energy cut-off of 400~eV and a smaller $k-$point grid of $4\times4\times4$ are used for the magnetic unit cell.
For the precise energy convergence, we make sure the energy difference between two consecutive runs is smaller than $10^{-5}$eV.

For energetics and band structure calculations of $M$Nb$_3$S$_6$, we use DFT by adopting the VASP code.
For the Berry curvature and AHC calculations in a dense $k-$mesh, we construct the spin-resolved real-space Hamiltonian by interpolating band structures obtained from first-principles:
\begin{equation}
\hat{H}=\sum_{\alpha\beta\sigma, \mathbf{R},\mathbf{R'}} t_{\alpha\beta}(\mathbf{R}- \mathbf{R'})\hat{c}^{\dagger}_{\mathbf{R}\alpha\sigma}\hat{c}_{\mathbf{R'}\beta\sigma}+ \sum_{\mathbf{r}_i}\hat{H}_U(\mathbf{r}_i),
\label{eq:Ham}
\end{equation}
where $t_{\alpha\beta}(\mathbf{R}-\mathbf{R'})$ is the spin-independent hopping integral between the Wannier function $\alpha$  within the  unit cell  $\mathbf{R}$ and the Wannier function $\beta$ within the  unit cell at $\mathbf{R'}$. 
The $t_{\alpha\beta}(\mathbf{R}-\mathbf{R'})$ parameters are calculated from the non-magnetic Kohn-Sham Hamiltonian in the basis of the maximally localized Wannier functions obtained using the Wannier90 code~\cite{Wannier90_2012,Pizzi2020}. 

For the spin-dependent part of the Hamiltonian, we adopt  multi-orbital Hubbard interaction correction term $\hat{H}_U$ for $d$ orbitals of every $M$ ion located at site $\mathbf{r}_i$:
\begin{eqnarray}
\hat{H}_U(\mathbf{r}_i) &=& \sum_{\alpha\beta\gamma\delta}\sum_{\sigma\sigma'} U_{\alpha\beta\gamma\delta} \hat{c}^{\dagger}_{\mathbf{r}_i\alpha\sigma}\hat{c}^{\dagger}_{\mathbf{r}_i\beta\sigma'}\hat{c}_{\mathbf{r}_i\gamma\sigma'}\hat{c}_{\mathbf{r}_i\delta\sigma} \nonumber \\ 
&\simeq& \sum_{\alpha\beta} (\bar{U}_H \: \hat{n}^i_{\alpha}\hat{n}^i_{\beta}+\bar{U}_F \: \hat{\mathbf{S}}^i_{\alpha}\cdot\hat{\mathbf{S}}^i_{\beta})
\label{eq:U}
\end{eqnarray}
where $U_{\alpha\beta\gamma\delta}$ is the on-site Coulomb interaction matrix between orbitals ($\alpha,\beta,\gamma,\delta$) centered at the magnetic site $\mathbf{r}_i$, $\bar{U}_H$ is the direct Coulomb interaction parameter for the density-density type interaction, $\bar{U}_F$ is the Hund's coupling parameter for the spin-spin type interaction,
$\hat{n}^i_\alpha$ is the density operator for the orbital $\alpha$ of magnetic atom $i$, and  $\hat{\mathbf{S}}^i_{\alpha}=\sum_{\sigma\sigma'}\hat{c}^{\dagger}_{\mathbf{r}_i \alpha\sigma}\mathbf{\sigma}_{\sigma\sigma'}\hat{c}_{\mathbf{r}_i\alpha\sigma'}$ is the local  spin operator.
Both $\bar{U}_H$ and $\bar{U}_F$ are taken averaged over orbital indices at the local site.
In the mean-field approximation, $\hat{H}_U$ can be written in terms of the one-particle Hamiltonian,
\begin{equation}
\hat{H}_U(\mathbf{r}_i) \simeq \sum_{\alpha}\sum_{\sigma\sigma'} \left(V^i\cdot\delta_{\sigma\sigma'}+\mathbf{J}^i\cdot\mathbf{\sigma}_{\sigma\sigma'}\right)\hat{c}^{\dagger}_{\mathbf{r}_i\alpha\sigma}\hat{c}_{\mathbf{r}_i \alpha\sigma'}
\label{eq:U2}
\end{equation}
where $V^i\simeq \sum_{\beta}\bar{U}_H\langle\hat{n}^i_{\beta}\rangle$ and $\mathbf{J}^i\simeq \sum_{\beta}\bar{U}_F \langle\hat{\mathbf{S}}^i_{\beta}\rangle$.
The local spin exchange potential $\mathbf{J}^i$   induces a spin polarization at the atomic site $i$ of the $M$ ion; within mean field, it is proportional to the local ordered magnetic moment itself.

For the magnetic states that we consider,  symmetry dictates that  $V^i$ is  site-independent ($V_i = V$), and $\mathbf{J}^i$ is site-dependent, but has constant magnitude, i.e., $|\mathbf{J}^i|=J$. Since both  $V$ and $J$ may differ for  different $M$ ions,  we determine them by fitting the band structure of Eqs.~(\ref{eq:Ham}- \ref{eq:U2}) to the one obtained with the spin-resolved DFT 
using the PBE exchange-correlation functional without spin-orbit coupling (see Fig.~\ref{fig:PM_band}), assuming the same magnetically ordered states. This allows us to construct a compact representation of band structure  that accounts for different stable magnetic orderings obtained within DFT. This band structure is  used  to efficiently calculate the AHC.
In addition to needing a quantitatively accurate and compact band structure, the $V$ and $J$ terms in the tight-binding Hamiltonian can parametrize the Coulomb interaction effects of \textit{spin-polarized} DFT bands starting with \textit{non-spin-polarized} DFT bands. We find that both interaction ($V$ and $J$) terms are necessary to capture the exact spin-polarized DFT bands although they can not treat orbital-dependent interactions due to the approximation used in Eq.$\:$\ref{eq:U2}.
Although most calculations in this paper are performed using spin-polarized DFT without tuning interaction parameters, we check the correlation effect on the AHC calculation by tuning the $J$ parameter as shown in Fig.$\:$\ref{fig:AHC_vs_J}.

The AHC $\sigma_{xy}$ is computed using the Berry curvature obtained from band structure as follows:
\begin{eqnarray}
\sigma_{xy}&=&\frac{e^2}{h}\frac{1}{2\pi}\sum_{n}\int d\mathbf{k} \:\:n_F(\epsilon_{n\mathbf{k}})\cdot\Omega_n^z(\mathbf{k})
\label{eq:sigma}
\end{eqnarray}
where $\Omega_n^z(\mathbf{k})=-Im\langle\partial_x u_{n\mathbf{k}}|\partial_y u_{n\mathbf{k}}\rangle$ is the Berry curvature of band $n$ with momentum $\mathbf{k}$ and $n_F$ is the Fermi function. $u_{n\mathbf{k}}$ is the periodic part of the Bloch wavefunction obtained by solving the spin-resolved Hamiltonian in Eq.~(\ref{eq:Ham}).
The calculation of Eq.~(\ref{eq:sigma}) requires  integration over a very fine $k-$mesh as an important large contribution of $\Omega_n^z$ can occur in a small region of the Brillouin zone (B.Z.).
Here, we adopt the Wannier-berry package~\cite{Wannier_berry} using the recursive adaptive refinement method based on symmetries for the smooth convergence of the $k-$mesh integration.
For the AHC calculations of $M$Nb$_3$S$_6$ compounds in the non-collinear spin configurations, we used the $10\times10\times8$ $k-$mesh with 10 recursive refinement iterations and temperature smoothing of the Fermi functions at 10~K which is well below the experimental Neel temperature of 29~K in the case of $M=$ Co.


\section{Results and Discussions}
\label{sec:Res}

\subsection{Magnetic structures and moments in $M$Nb$_3$S$_6$}
\label{sec:magnetism}

In $M$Nb$_3$S$_6$ compounds, transition metal $M$ ions are intercalated between NbS$_2$ layers,  forming triangular lattices stacked  along the $c-$axis at two  distinct Nb sites locations in an alternating fashion (see Fig$\:$\ref{fig:struct}a,b).
The issue of the magnetic ground-state in $M$Nb$_3$S$_6$ has not been settled yet experimentally, 
and we therefore consider a variety of collinear and non-collinear spin configurations of the $M$ ions (see Fig$\:$\ref{fig:struct}d-k). Some of them have the magnetic unit cell identical to the primitive unit cell (d-e); some double unit cell (f-g), and some -- quadruple (h-k). 
The magnetic moment and the energy for each compound and spin structure are computed using DFT.

All experimentally relevant  spin structures can be obtained from the following  $ansatz$ for the  three-dimensional spin vector $\mathbf{S}$:
\begin{equation}
 \mathbf{S}(\mathbf{r})=\left(A\cos{\mathbf{q_1}\cdot\mathbf{r}}, B\cos{\mathbf{q_2}\cdot\mathbf{r}}, C\cos{\mathbf{q_3}\cdot\mathbf{r}}\right),  
 \label{eq:ABC}
\end{equation}
where $A, B, C$ are constants and $\mathbf{q_1}, \mathbf{q_2}, \mathbf{q_3}$ are  modulation wavevectors of the orderings. Note that for $\mathbf{q_i}$ that are either reciprocal or half of the reciprocal lattice vectors of the primitive unit cell, the magnitude of spins is the same for all sites, $|\mathbf{S}(\mathbf{r})|=S=\sqrt{A^2 + B^2 + C^2} $.

The simplest collinear cases are c-FM in which all $M$ spins are aligned ferromagnetically  (see Fig.~\ref{fig:struct}d) and c-AFM0 where $M$ spin directions are FM in the $a$-$b$ plane but anti-parallel between two crystallographically inequivalent layers (see Fig.~\ref{fig:struct}e). 
In both cases, the magnetic unit cell coincides with the primitive unit cell -- all $q_i = 0$. 
We note that in the crystal structure of $M$Nb$_3$S$_6$, the inversion symmetry is broken; this allows for nonzero Dzyaloshinskii-Moriya interaction that tends to favor in-plane alignment of spins \cite{OllePaper}. 

In c-AFM1 and c-AFM2 cases, spins are ferromagnetically ordered along one of the in-plane Co-Co bond directions and modulated antiferromagnetically along the other in-plane Co-Co bond directions. 
In this so-called $1q$ structure, the ordering wavevector is half of the in-plane reciprocal vector (a high-symmetry point $M$ in the primitive Brillouin zone), which is orthogonal to the ferromagnetically ordered direction.
In the $1q$ structure, there are two distinct relative spin arrangements of $M$ spins  in the adjacent layers, one favors AF interlayer coupling (see Fig.~\ref{fig:struct}f) and the other FM interlayer coupling (see Fig.~\ref{fig:struct}g).

In the non-collinear cases, we consider the magnetic unit cell containing four distinct spin orientations per layer (see Fig.~\ref{fig:struct}b dashed line) 
These orders involve two  or three ordering wave vectors that correspond to $M$ points in the primitive Brillouin zone. 
The coplanar ordering  (cpl-) involves two ordering vectors (``2$q$" structure) and spins that lie in a single plane in the spin space; their sum is zero, so they can be arranged into a rhombus (see Fig.~\ref{fig:struct}h-i).

The non-coplanar (ncpl-) AFM  involves all three $M$-point wave-vectors (``3$q$" structure). The spin orientations correspond to the directions from the center toward vertices of a regular tetrahedron~\cite{martin_scalar_2008}. 
The 3$q$ state on a triangular lattice is special as it corresponds to a  scalar chirality, $\chi_{ijk} = {\bf S}_i\cdot [{\bf S}_j\times {\bf S}_k]$, that is constant for all elementary triangular plaquettes $(i,j,k)$. For two dimensional systems this leads to Anomalous Hall effect even in the absence of spin-orbit interactions \cite{martin_scalar_2008, kato_stability_2010}. In the bilayer structure of $M$Nb$_3$S$_6$, the  two $M$ layers can have either the same or the opposite scalar chirality.  Only in the former case we anticipate AHC to be present, since in the latter case layer translation followed by time reversal is a symmetry that prohibits finite AHC. In our DFT calculations we compute the energies of both states (ncpl-AFM1, see Fig.~\ref{fig:struct}j and ncpl-AFM2, see Fig.~\ref{fig:struct}k).

We finally note that all the AF states that we consider can be smoothly distorted into each other: $3q$ state can be ``flattened" into a $2q$ state by decreasing, e.g., the $z$ component of spin (coefficient $C$ in Eq. (\ref{eq:ABC})); further, the $2q$ structure can be distorted into $1q$ by reducing one of the remaining spin components, e.g. $y$ (coefficient $B$), to zero. 
In this work we do not exhaust all the possible ordered states. Instead, we compare the most  symmetric states, where nonzero amplitudes $A,B,C$ are all equal in magnitude.

\begin{table}[ht]
    \centering
    \begin{tabular}{|c|c|c|c|c|} 
    \hline
      Mag. mom. \& energy & $M$=Mn & $M$=Fe & $M$=Co & $M$=Ni \\ [1.0ex] 
    \hline
      $m/m_{0}$ [$\mu_B$] & 3.9/5.0 & 3.1/4.0 & 1.5/3.0 & 0.7/2.0 \\ 
      \hline
      c-FM [eV] & {\bf -74.900} & -73.412 & -72.211 & -69.734 \\ 
      c-AFM0 [eV] & -74.869 & -73.423 & -72.221 & -69.743 \\ 
      c-AFM1 [eV] & -74.814 & -73.425 & -72.242 & -70.576 \\
      c-AFM2 [eV] & -74.811 & -73.404 & -72.222 & -70.588 \\
      cpl-AFM1 [eV] & -74.816 & -73.423 & -72.244 & -70.576 \\
      cpl-AFM2 [eV] & -74.811 & -73.410 & -72.223 & -70.589 \\
      ncpl-AFM1 [eV] & -74.816 & {\bf -73.426} & {\bf -72.245} & -70.576 \\
      ncpl-AFM2 [eV] & -74.811 & -73.408 & -72.224 & {\bf -70.589} \\
     \hline
    \end{tabular}
    \caption{
Spin magnetic moments ($m$) of the $M^{2+}$ ions 
and total energies per formula unit of $M$Nb$_3$S$_6$ comparing collinear (c-) FM \& AFM states and non-collinear coplanar (cpl-) \& non-coplanar (ncpl-) spin states computed using DFT. 
The ordered magnetic moments are almost the same for all magnetic states; the spin magnetic moments ($m_0$) for the free $M^{2+}$ ions are also given.
Bolded values correspond to the lowest energy states.  
}
\label{tbl:energy2}
\end{table}

Table\:\ref{tbl:energy2} shows spin magnetic moments and total energies per formula unit of the c-FM, c-AFM, cpl-AFM, and ncpl-AFM spin structures in $M$Nb$_3$S$_6$ with $M=$Mn, Fe, Co, and Ni computed using DFT.
The spin magnetic moments are calculated by integrating the spin density over an atomic sphere given by the VASP code.
The trend of calculated spin magnetic moments in all compounds can be understood from the high-spin configurations in the divalent transition metal $M^{2+}$ ions, consistently with the value obtained from neutron scattering experiment in the case of $M=$Co~\cite{parkin_magnetic_1983}. 
The spin magnetic moments of high-spin states in free $M^{2+}$ ions, denoted as $m_0 \:(=g\cdot S_z)$ in Table\:\ref{tbl:energy2}, are 5.0$\mu_B$ ($M$=Mn), 4.0$\mu_B$ ($M$=Fe), 3.0$\mu_B$ ($M$=Co), and 2.0$\mu_B$ ($M$=Ni) as the number of unpaired electrons changes from 5 to 2 (from Mn to Ni).
The calculated spin magnetic moments in $M$Nb$_3$S$_6$ in Table\:\ref{tbl:energy2},  are rather reduced from these free ion results, e.g., $S=$1.5$\mu_B$ for $M=$Co. Our calculated spin magnetic moments are almost unchanged regardless of the spin configurations, namely whether they are collinear or noncollinear.
The experimentally measured value of the spin moment in CoNb$_3$S$_6$ is 2.73$\mu_B$~\cite{parkin_magnetic_1983} which is  smaller than the free ion value although somewhat larger than the DFT result.
The reduction of spin moment in $M$Nb$_3$S$_6$ compared to the free ion value is expected since the local  moment is coupled to itinerant Nb bands and can be screened relative to the ionic value. 
The further reduction of the moments obtained in DFT compared to experiment can be due to the underestimation of correlations in DFT as the on-site interaction $U$ effect is not fully captured for transition metal ions. 
We also evaluated the role of spin-orbit coupling in these compounds by comparing the collinear magnetic energies along various ordered spin directions. Regardless of the spin directions, energies of all compounds are almost the same (within 1meV) except FeNb$_3$S$_6$ for which the energy of spin direction along the easy plane was lower than one along the $z-$axis by $\sim$18meV per formula unit. Given the small difference of $\sim$1meV between collinear and non-collinear spin states, it is important to include the spin-orbit effect for the energy calculation in FeNb$_3$S$_6$. The spin-orbit effect is much smaller for other compounds, which is typically expected in these transition metal ions with 3$d$ orbitals.

All $M$Nb$_3$S$_6$ compounds studied here favor the ncpl-AFM structures energetically except the $M= $~Mn case (see Table \ref{tbl:energy2}).
We find that the variant with Mn orders ferromagnetically. Qualitatively, this is consistent with the Mn magnetic moment being the largest, and thus most strongly coupled to itinerant electrons; this favors ferromagnetic ordering via the conventional double-exchange mechanism. 
Indeed, the magnetic ground state exhibits metallic behavior with large spins of the  intercalated Mn ions  strongly hybridized with itinerant Nb $d_{z^2}$ orbitals in the NbS$_2$ layers. Due to strong hybridization, any misalignment of ordered moments will frustrate electronic kinetic energy, and therefore is energetically penalized. The next energy state is c-AFM0, with ferromagnetic Mn planes stacked antiferromagnetically along the $c$ axis. This is consistent with having dominant ferromagnetic coupling between Mn in individual layers, and a weaker ferromagnetic coupling between them.

On the other hand, for smaller moments and thus weaker exchange fields, the physics is expected to be controlled by Fermi surface instabilities. 
Given the hexagonal symmetry of the crystal, having a magnetic order with multiple ordering wave vectors then becomes favorable as it allows us to gap  larger sections of the Fermi surface. The $3q$ ncpl-AFM state is particularly attractive: At a commensurate carrier density, such weak-to-intermediate coupling instability can fully gap the  Fermi surface, producing an insulator with quantized AHC \cite{martin_scalar_2008, akagi2010}. 
Indeed, the lowest energy spin configuration that we were able to obtain for 
$M$Nb$_3$S$_6$ for $M$=Fe, Co, and Ni correspond to the $3q$ state; it is ncpl-AFM1 for $M$=Fe, Co and ncpl-AFM2 for $M$=Ni. The difference between these noncoplanar states is that one has the same sign of scalar spin chirality in both $M$ layers within the unit cell (ncpl-AFM1), while the other (ncpl-AFM2)  has  opposite  in sign chiralities in two layers.  Because of the scalar chirality structure, we expect  non-zero net AHC for Fe and Co cases, and zero AHC for Ni case. This is indeed confirmed numerically in the later subsection (see Fig.~\ref{fig:AHC}).

It is useful to also examine the  DFT results in Table \ref{tbl:energy2} for other states besides the lowest energy ones. From the magnetic structure, Fig.~\ref{fig:struct}, the (c,cpl,ncpl)-AFM1 are favored by the antiferromagnetic  nearest neighbor interlayer coupling, while (c,cpl,ncpl)-AFM2 are by the  ferromagnetic one.
The pattern of energies is consistent with this coupling determining the type of the ordered states: For a given material, all x-AFM1 states are consisently above or below the corresponding x-AFM2.
We also notice  that coplanar and non-coplanar states are extremely close in energy, so it is conceivable that in the real materials the energy balance  could be tipped in the opposite direction.

\subsection{Electronic structure of $M$Nb$_3$S$_6$}
\label{sec:bands}

The DFT calculations described above provide us with the candidate magnetic ground states, and we anticipate that ncpl-AF1 is likely to have significant AHC. In order to compute the Hall conductivity we need both the wave functions and energies of electrons with high momentum resolution. This is very computationally expensive within DFT, particularly so if the  magnetic unit cell is enlarged and   magnetization is noncollinear. 

\begin{figure*}[!ht]
\includegraphics[width=0.268\linewidth]{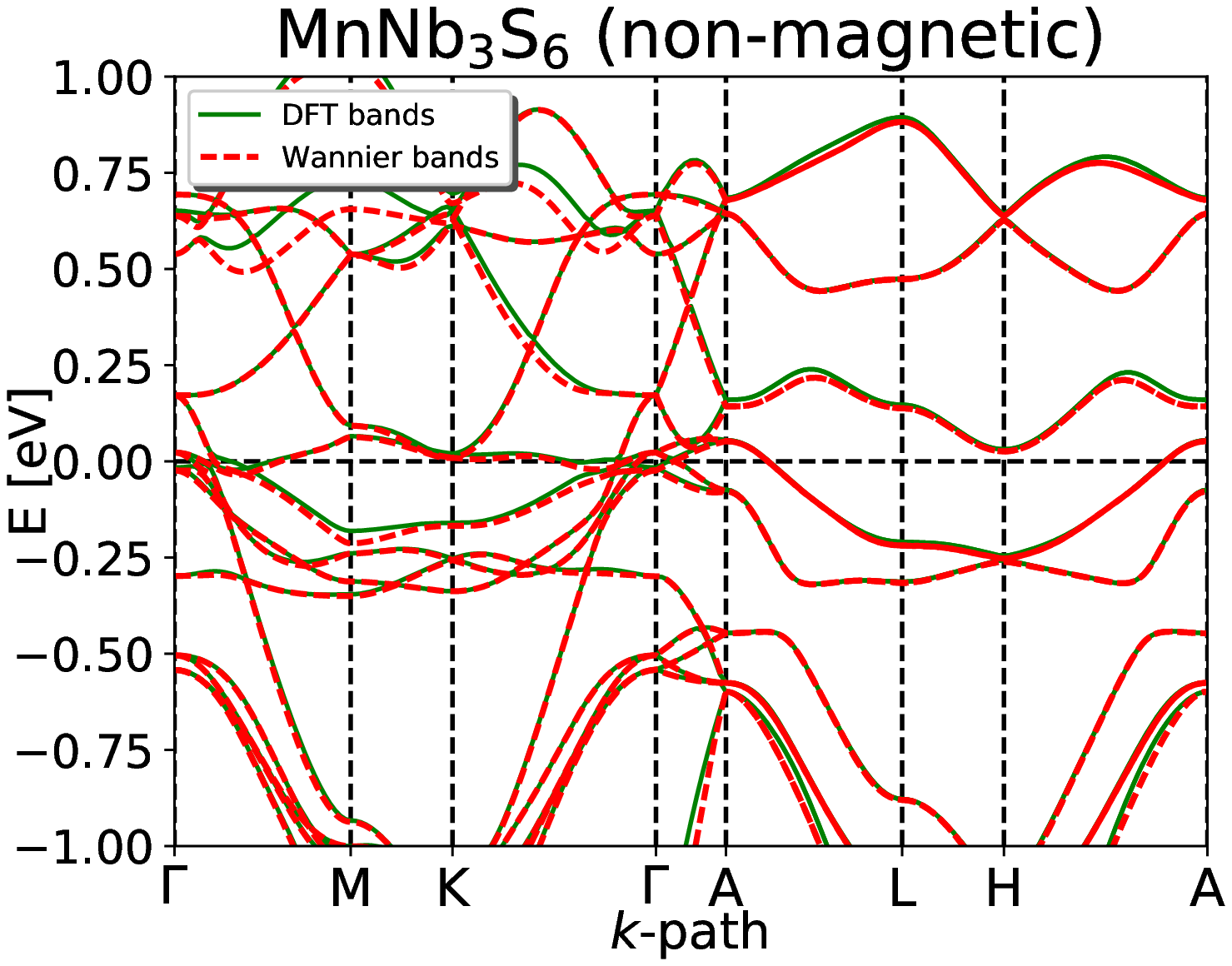}
\hspace{-0.65cm}
\includegraphics[width=0.268\linewidth]{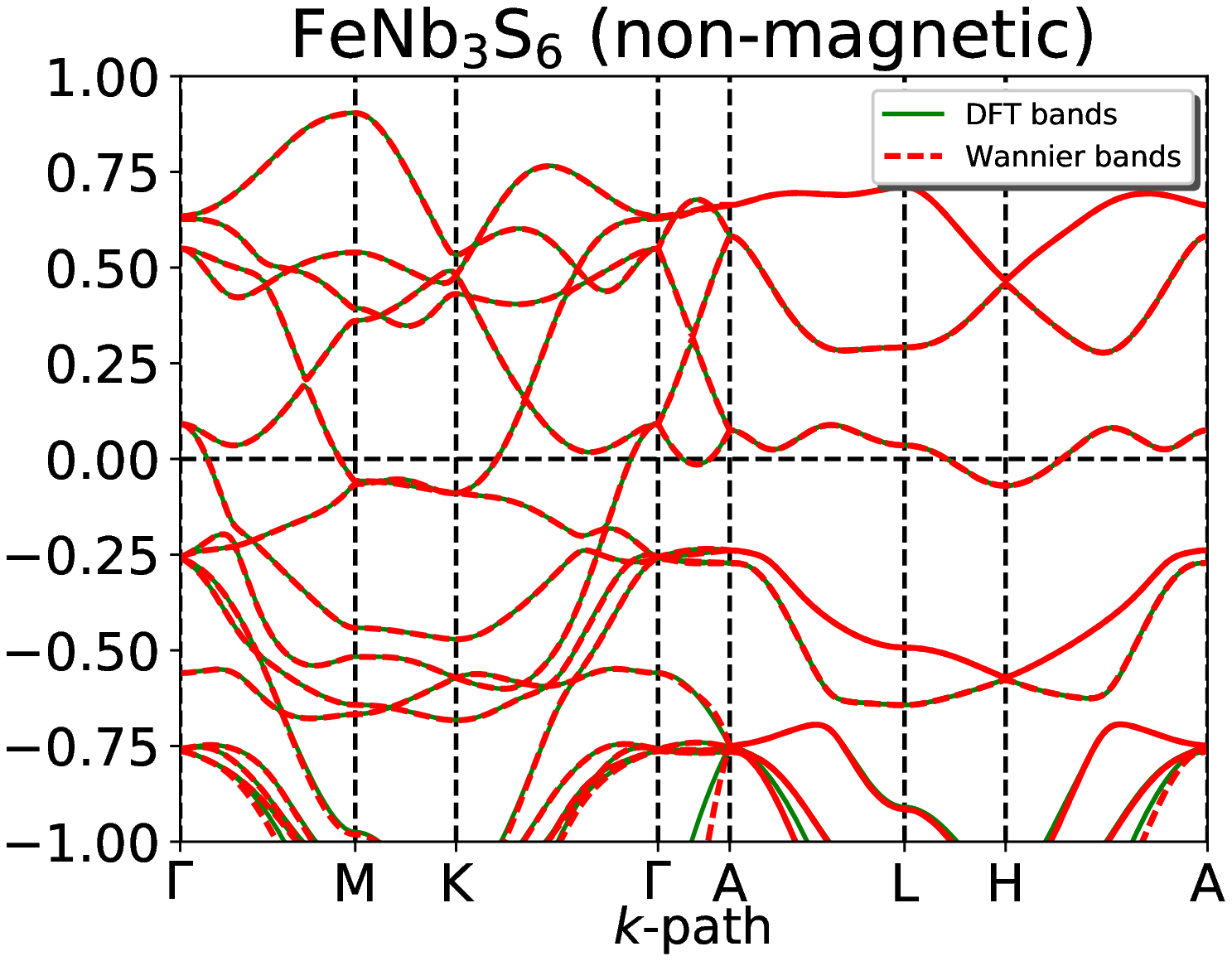}
\hspace{-0.65cm}
\includegraphics[width=0.268\linewidth]{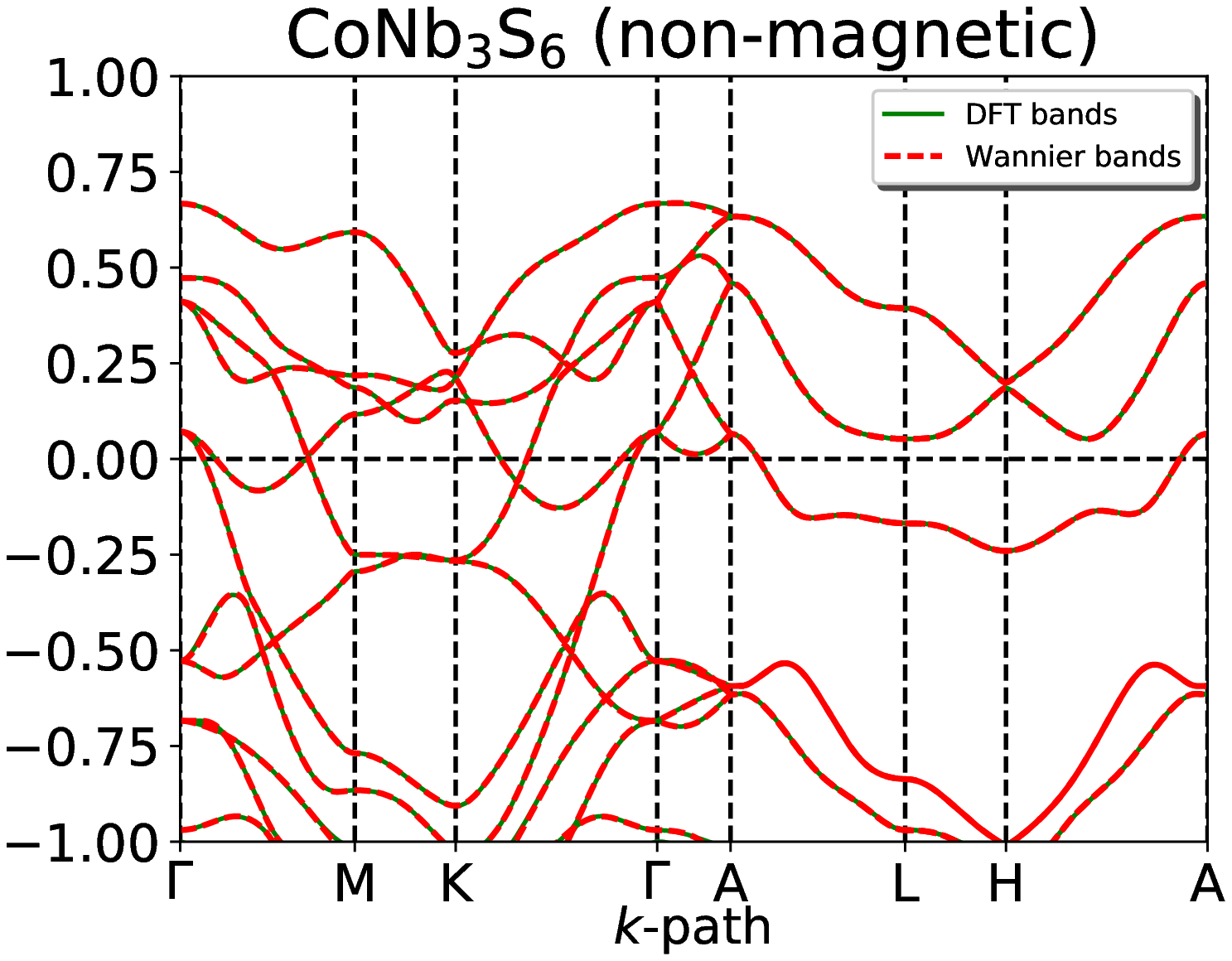}
\hspace{-0.65cm}
\includegraphics[width=0.268\linewidth]{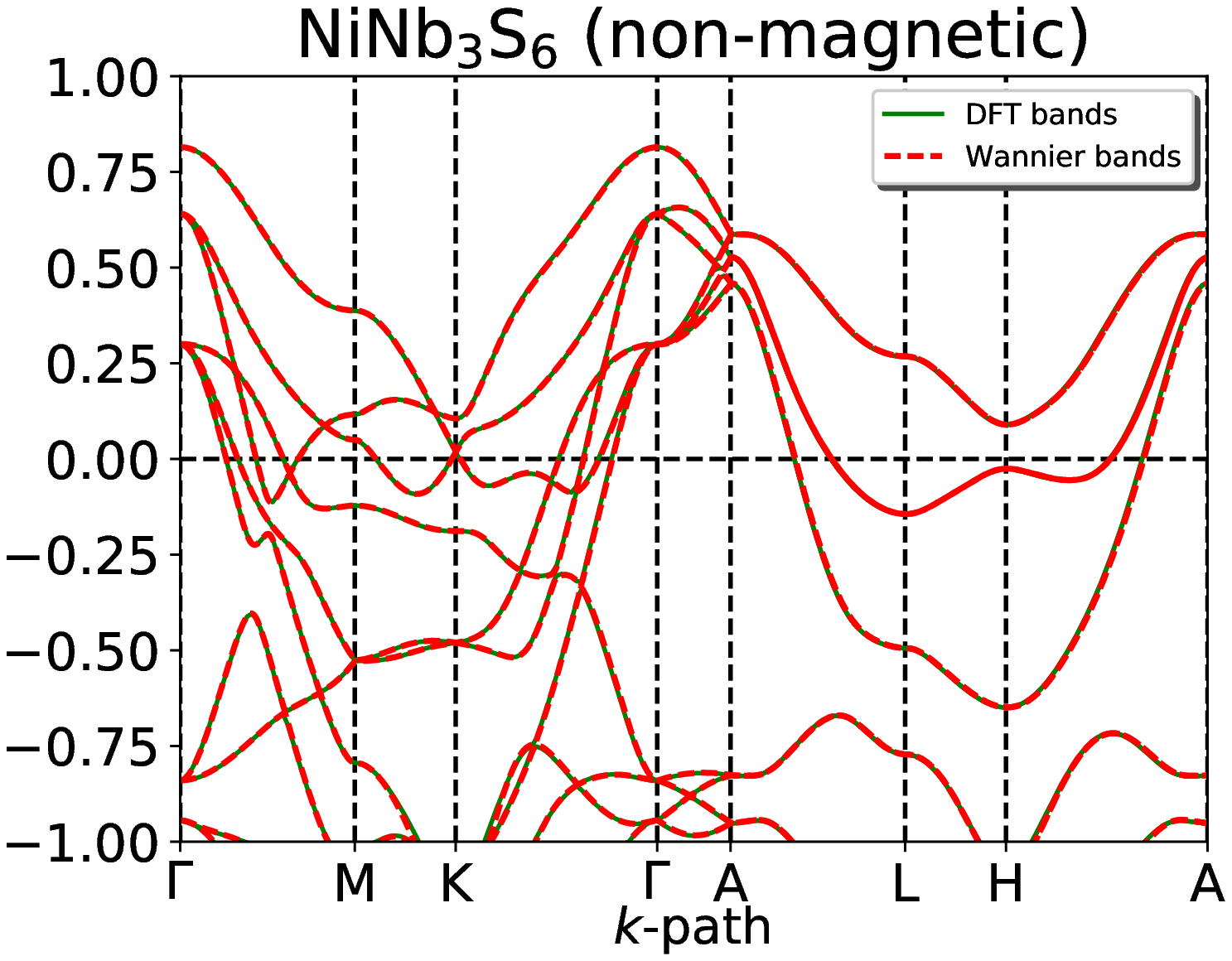}\\
\includegraphics[width=0.268\linewidth]{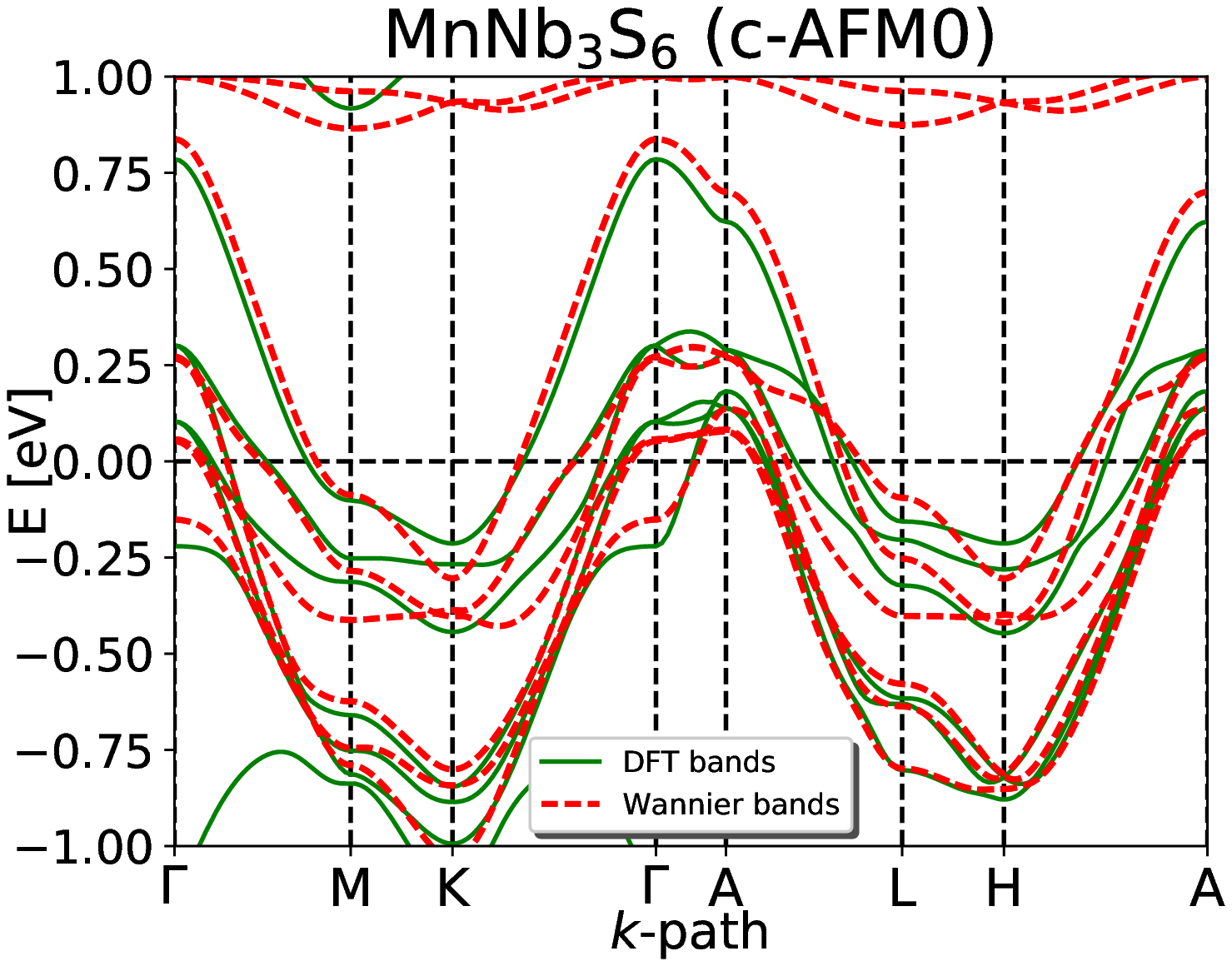}
\hspace{-0.65cm}
\includegraphics[width=0.268\linewidth]{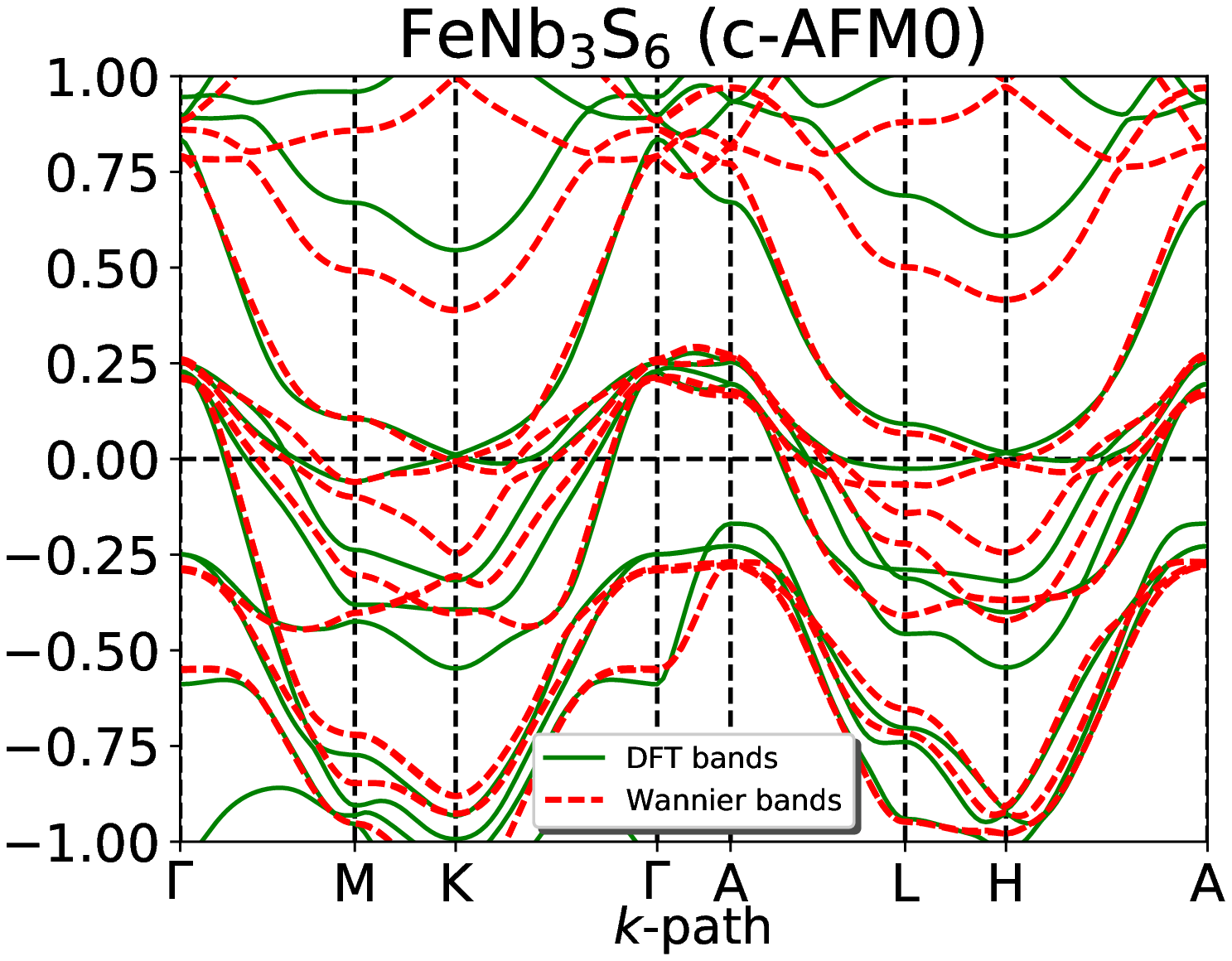}
\hspace{-0.65cm}
\includegraphics[width=0.268\linewidth]{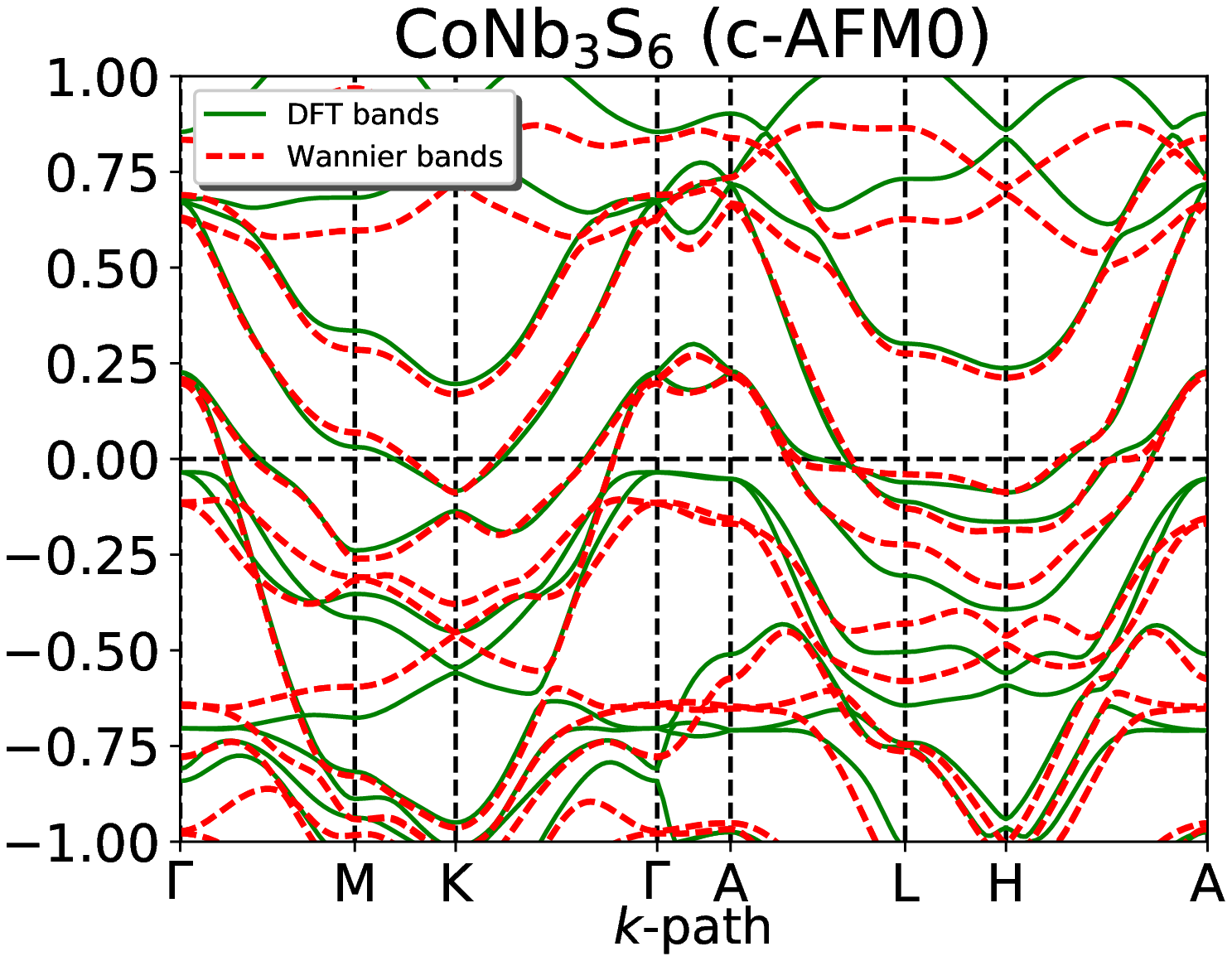}
\hspace{-0.65cm}
\includegraphics[width=0.268\linewidth]{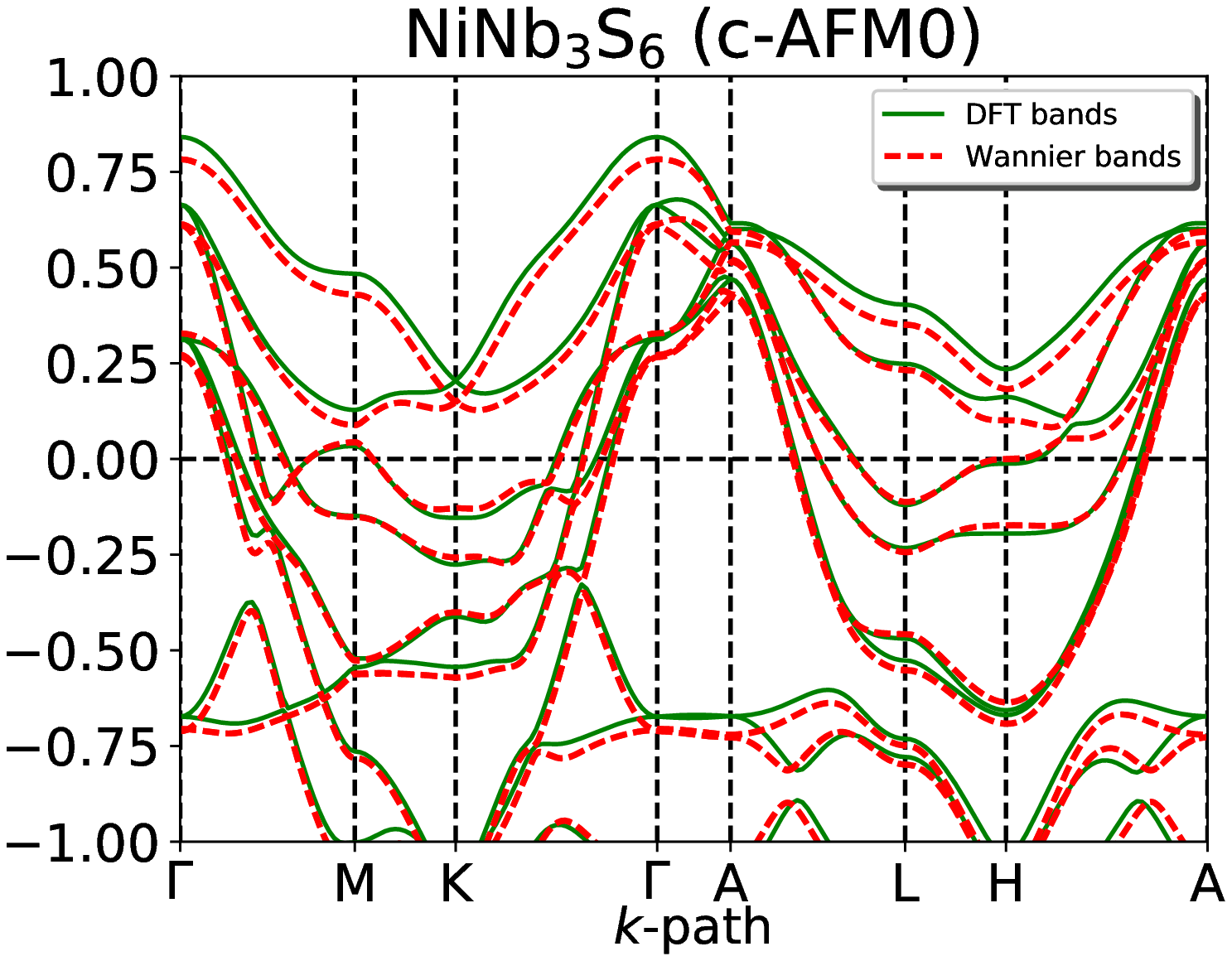}
\caption{Top row: Comparisons of non-magnetic band structure of $M$Nb$_3$S$_6$ with $M$=Mn, Fe, Co, and Ni obtained using DFT (solid line) and the Wannier interpolation (dashed line). Bottom row: Comparisons of collinear AFM (c-AFM0) band structures of $M$Nb$_3$S$_6$ with $M$=Mn, Fe, Co, and Ni obtained using DFT (solid line) and the Wannier Hamiltonian in Eq.~(\ref{eq:Ham})  with material-specific  $J$ and $V$ values (dashed line). The $k-$path is along the high-symmetry points in the B.Z. of the primitive unit cell (see Fig.~\ref{fig:struct}c red line).
}
\label{fig:PM_band}
\end{figure*}

To reduce the computational cost,  we construct  effective tight-binding models for the materials of interest by fitting the relevant bands  using Wannier interpolation technique~\cite{Wannier90_2012}. We do this in two steps.
First, we construct the real-space Hamiltonian (Eq.~(\ref{eq:Ham})) with the  hopping parameters ($\mathbf{t_{\alpha\beta}}$) between Wannier orbitals in the paramagnetic state obtained with the Wannier90 code~\cite{Pizzi2020}. As can be seen from the upper panels of Fig.~\ref{fig:PM_band}, the matching between the Wannier bands and the non-magnetic DFT bands is almost perfect.
We then treat  the local direct interaction ($V$) and the spin exchange ($\mathbf{J}$) terms  on the transition metal $M$ as free parameter to fit the band structure in magnetically ordered states. We use the c-AFM0 band structure computed without SOC since it is the simplest AFM structure that has the same unit cell as the crystal itself.

\begin{figure}[!ht]
\includegraphics[width=0.9\linewidth]{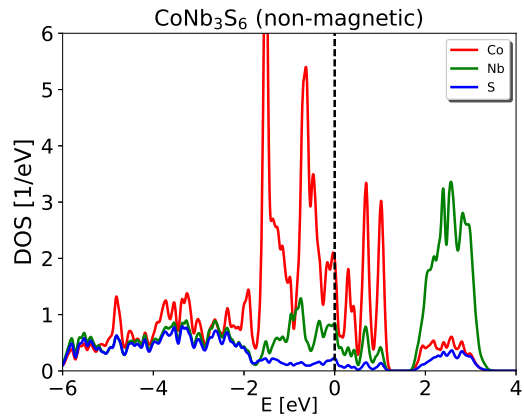}
\caption{Density of states in CoNb$_3$S$_6$ projected to Co, Nb, and S ions computed using non-magnetic DFT.}
\label{fig:PM_DOS}
\end{figure}

Fig.~\ref{fig:PM_DOS} shows the density of states for CoNb$_3$S$_6$ projected to different ions. The Co $d$ orbitals are concentrated near the Fermi energy in  the range between $-2~e$V and $1~e$V, and they are also strongly hybridized with the itinerant Nb $d_{z^2}$ orbital. Other Nb $d$ orbitals are located at higher energies, over 2~eV above the Fermi level. The hybridization with the S ions occurs at much lower energies, below $-2$~eV. For other transition metals $M$, the orbital arrangement  and energy scales are similar. Therefore, we  use Nb $d_{z^2}$  and $M$ ion's $d$ Wannier orbitals for constructing the non-magnetic Hamiltonian (Eq.~(\ref{eq:Ham}); $\hat{H}_U=0$).

Fig.~\ref{fig:PM_band}  top panel compares the obtained Wannier bands (red dashed lines) to non-magnetic DFT band results (green solid lines)
along the high-symmetry directions in the B.Z. of the primitive cell (see Fig.~\ref{fig:struct}c black line B.Z.).
The Wannier bands obtained from the non-magnetic Hamiltonian in Eq.~(\ref{eq:Ham})  reproduce the DFT bands almost exactly for all $M$Nb$_3$S$_6$ compounds. 
As the $M$ ion changes from Mn to Ni, the occupancy of $d$ orbitals increases from $d^5$ to $d^8$ and bands with $M$ ion character shift down in energy.
This change is particularly noticeable by following the band evolution near the $k_z=\pi$ plane in the momentum space ($A-L-H-A$ line in Fig.~\ref{fig:PM_band}).

To be able to construct the band structure of the magnetically ordered states,  we introduce $V$ and $\mathbf{J}$ parameters in Eq.~(\ref{eq:U2}). 
The $\mathbf{J}$ field is an effective spin-exchange potential, which is proportional to the magnetic moments of the $M$ site; it is therefore  site-dependent due to the varying spin directions in the AFM state. The $V$ parameter is the local effective potential for the density-density interaction and it determines the relative position of $M$ ion bands compared to the Nb $d_{z^2}$ band; it is site-independent.
Figure~\ref{fig:PM_band} bottom panels compare the spin-polarized band structure obtained from the Wannier Hamiltonian (dashed lines) to spin-resolved DFT (solid lines) calculations for the c-AFM0 state (collinear state with FM planes of $M$, stacked antiferromagnetically along the $c$ axis).
As seen in Fig.~\ref{fig:PM_band}, including $\mathbf J$ changes the band structure noticeably relative to the non-magnetic bands except for $M$=Ni. 
Focusing on the states nearest to the Fermi level, we find the following parameters that allow an accurate fit of Wannier bands to DFT ($J$  is $|\bf J|$, which is the same on all magnetic sites in the states that we consider): $J=1.3$~eV and $V=-0.4$~eV for Mn, $J=0.8$~eV and $V=-0.5$~eV for Fe, $J=0.5$~eV and $V=-0.1$~eV for Co, and $J=0.2$~eV and $V=0$~eV for Ni. Since the spin-exchange potential $J$ is the product of the Hund's coupling $\bar{U}_F$ and the magnitude of the ordered moment, and since the Hund's coupling does not vary much with ion $M$, the correlation between the value of $J$ and the size of the ordered moments is expected.  Indeed, Mn has the largest moment, and, predictably, has  the largest value of $J$.

\subsection{Results of AHC calculations}
\label{sec:AHC}

\begin{figure*}[!ht]
\includegraphics[width=0.269\linewidth]{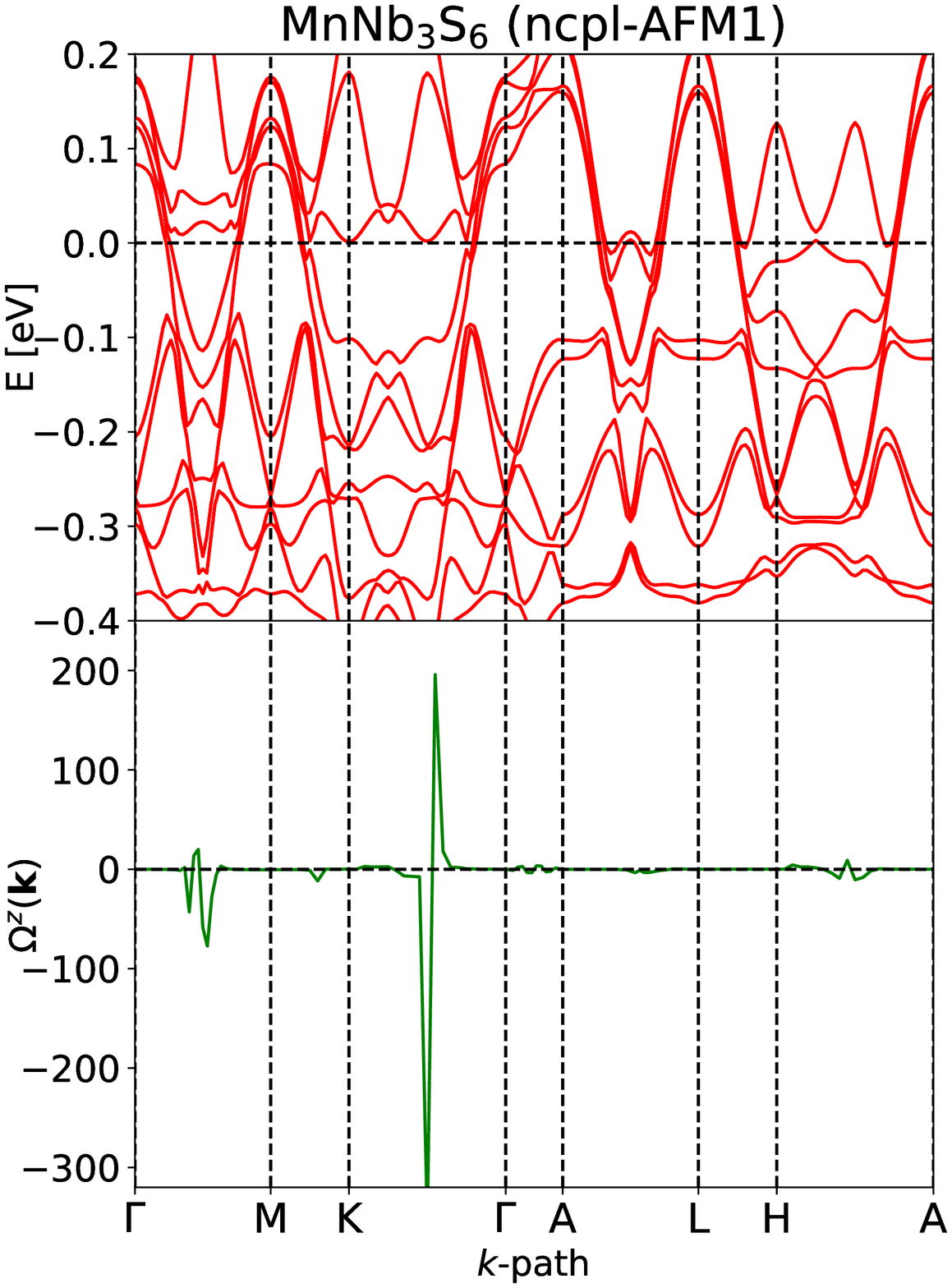}
\hspace{-0.645cm}
\includegraphics[width=0.269\linewidth]{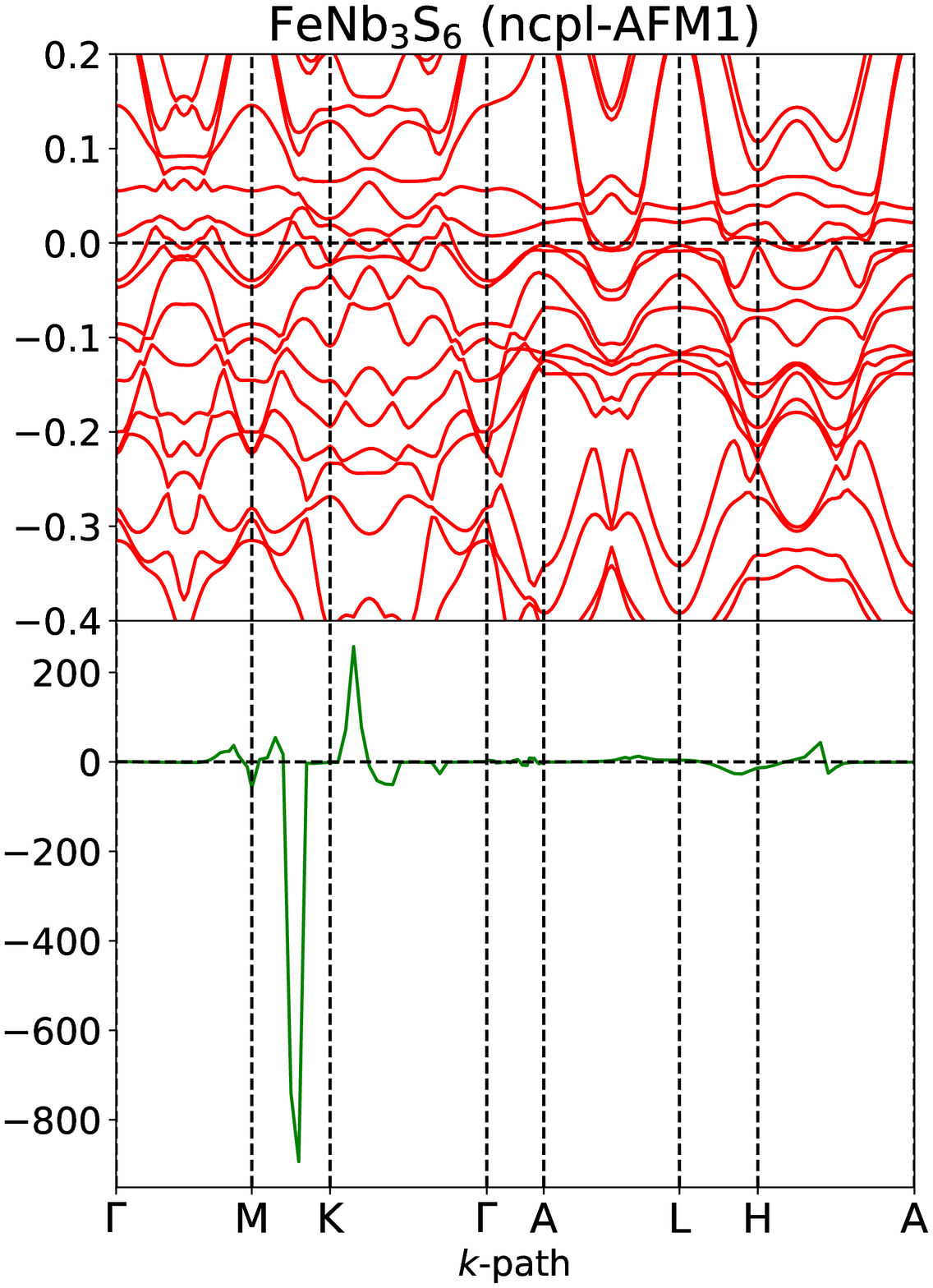}
\hspace{-0.645cm}
\includegraphics[width=0.269\linewidth]{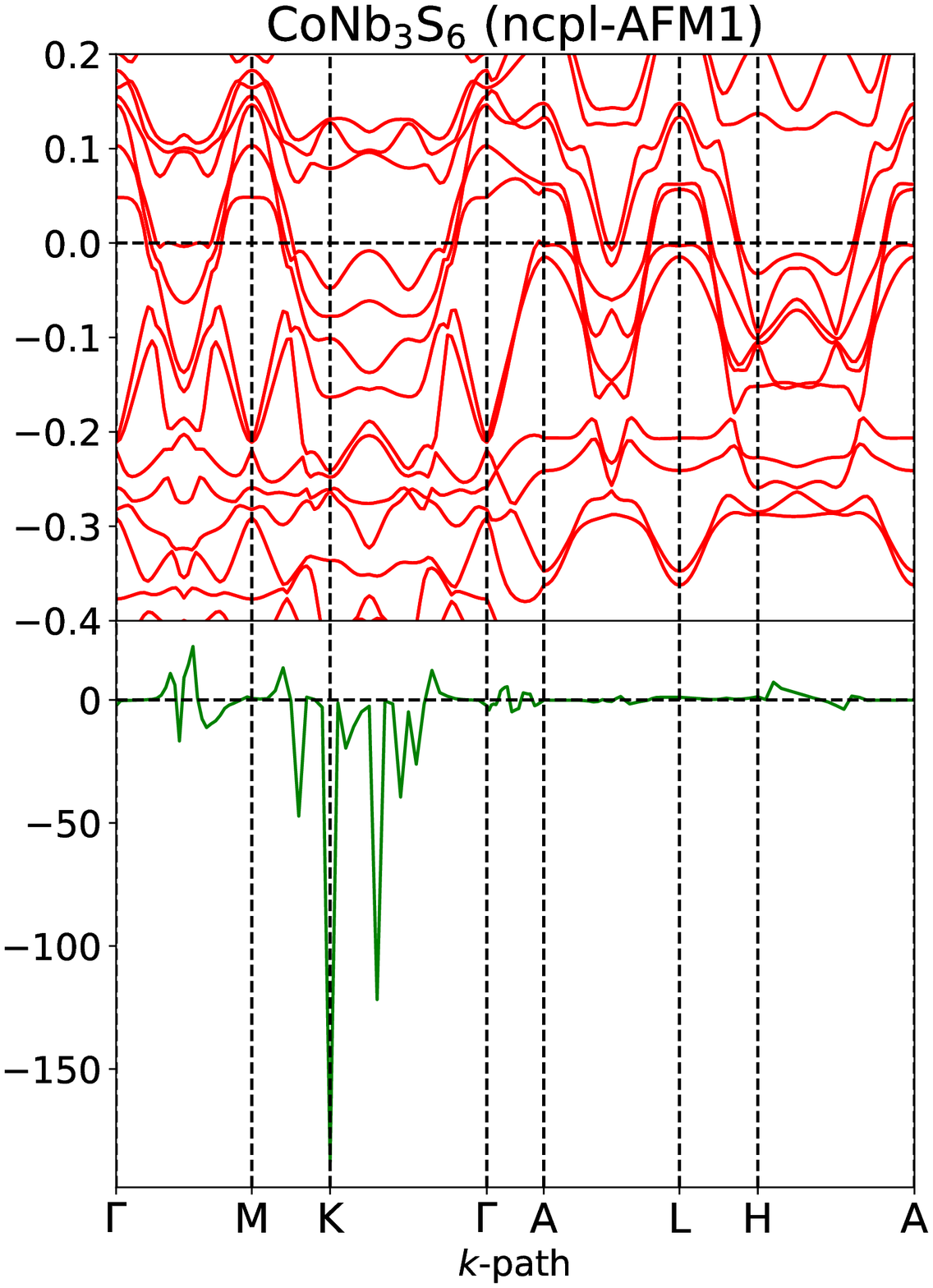}
\hspace{-0.645cm}
\includegraphics[width=0.269\linewidth]{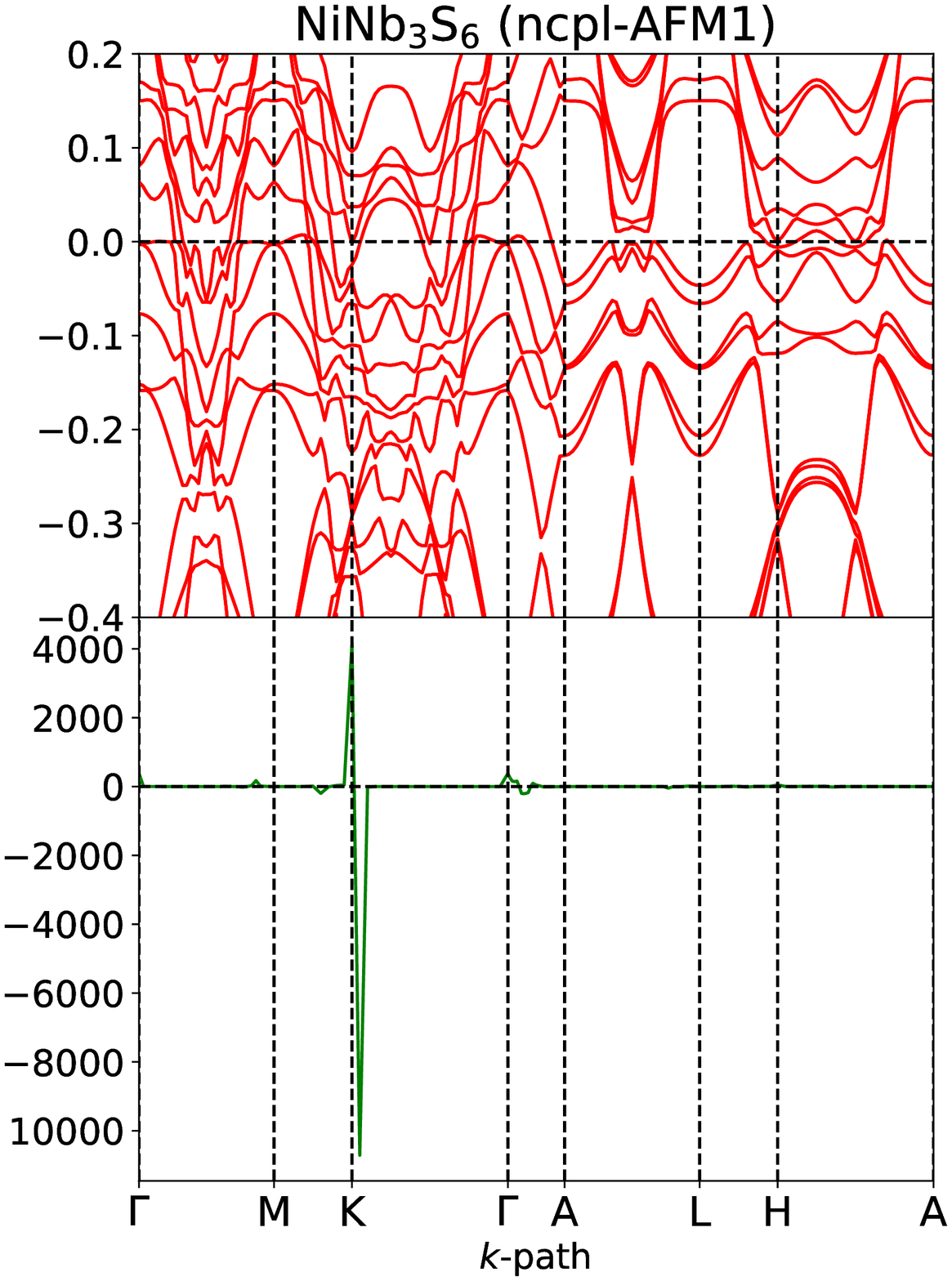}
\caption{Top panel: The band structure of the non-coplanar antiferromagnetic (ncpl-AFM1) spin structure in $M$Nb$_3$S$_6$ for $M$=Mn, Fe, Co, and Ni obtained from the Wannier Hamiltonian in Eq.~(\ref{eq:Ham}). Here, the same $V$ and $|\mathbf{J}|$ values are used for each material as the c-AFM0 case in Fig.~\ref{fig:PM_band}. The $k-$path is along the high-symmetry points in a smaller B.Z. of the magnetic unit cell (see Fig.~\ref{fig:struct}c for the definition of the smaller B.Z.). Bottom panel: The Berry curvature $\Omega(\mathbf{k})$ of $M$Nb$_3$S$_6$ for $M$=Mn, Fe, Co, and Ni summed over all occupied bands plotted along the same $k-$path as the top band structure.}
\label{fig:non_col_band}
\end{figure*}

\begin{figure}[!h]
\includegraphics[width=0.9\linewidth]{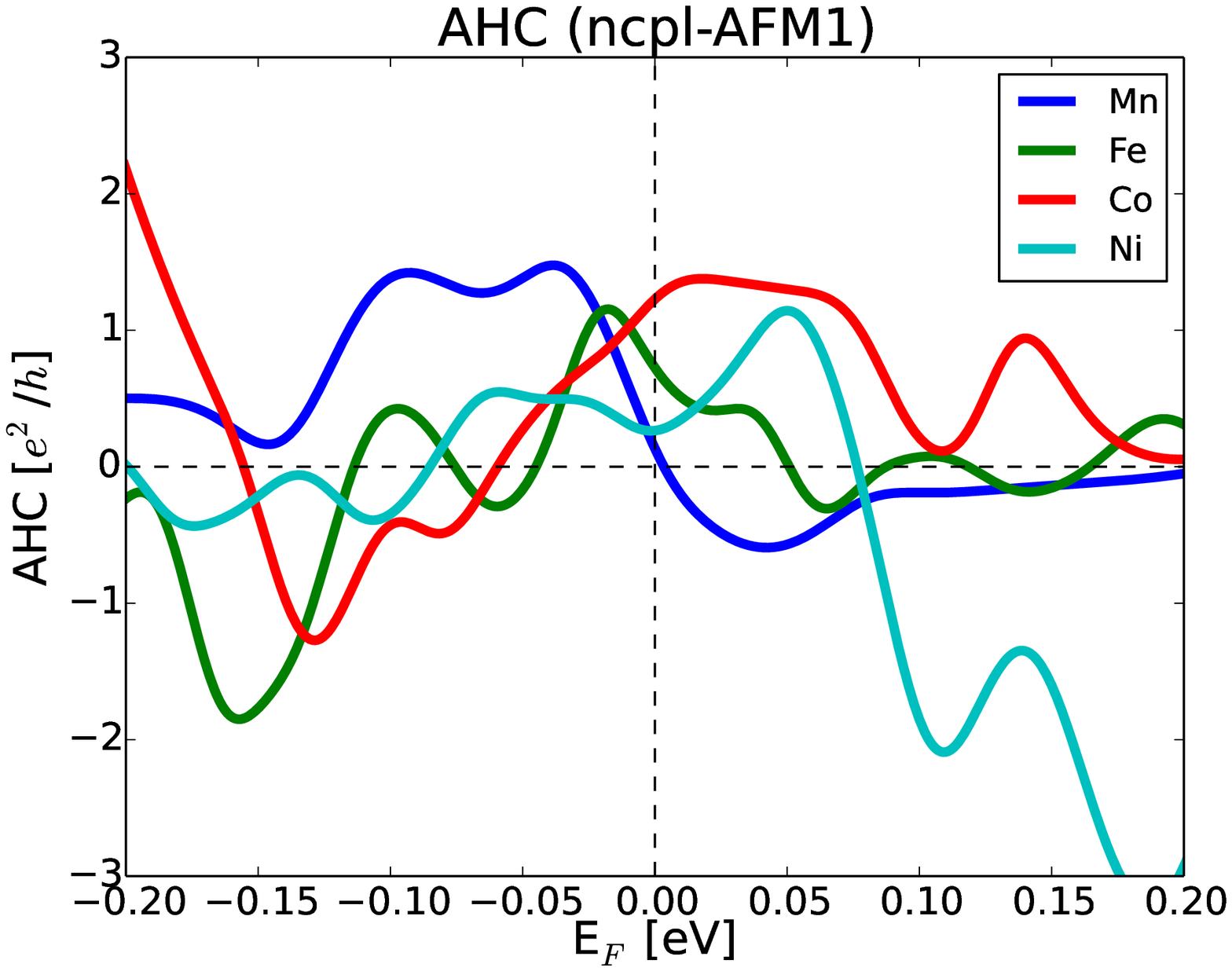}\\
\includegraphics[width=0.9\linewidth]{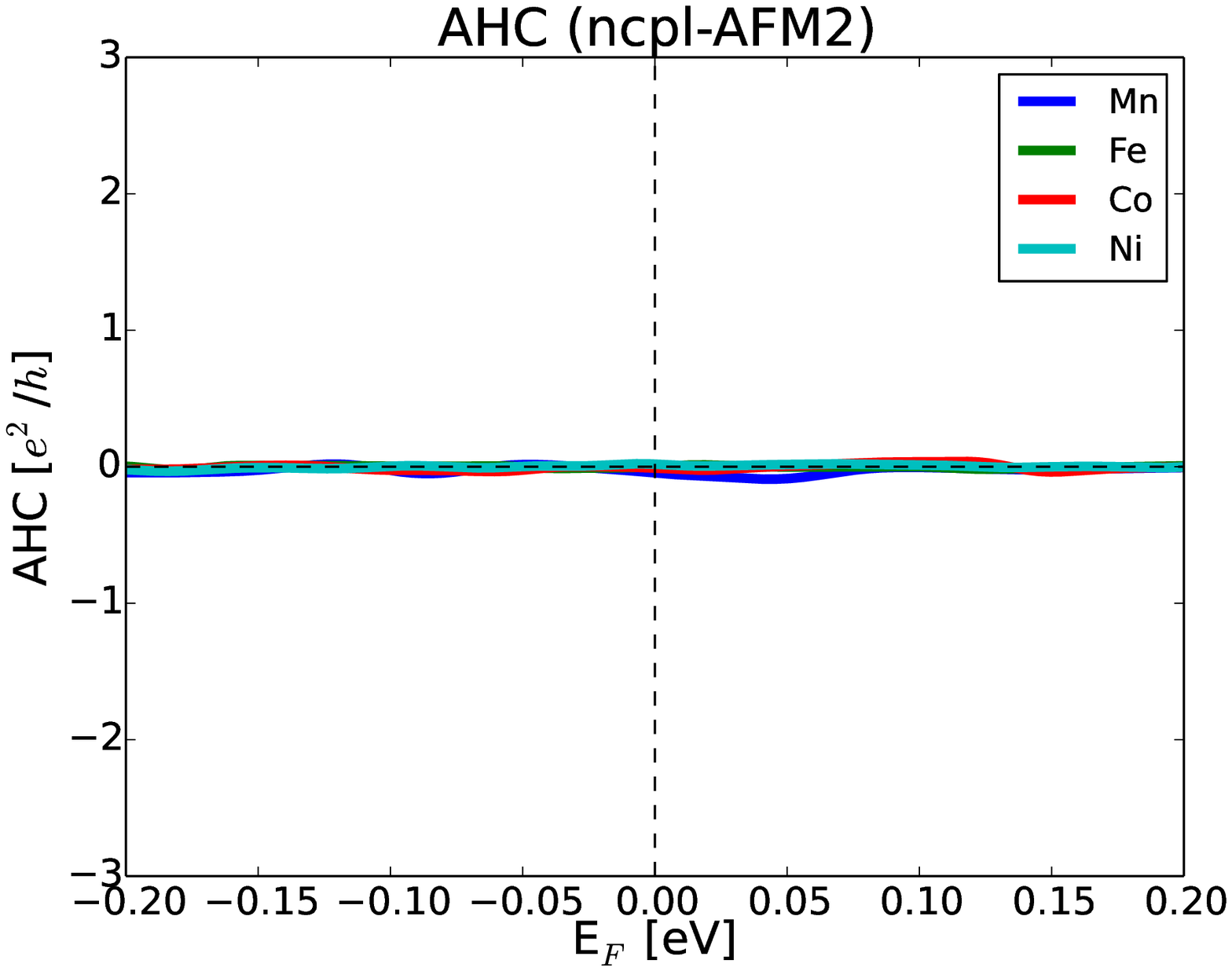}
\caption{Calculated AHC of $M$Nb$_3$S$_6$ as a function of the Fermi energy $E_F$ for both the ncpl-AFM1 (top panel) and the ncpl-AFM2 (bottom panel) spin structures for  $M$=Mn, Fe, Co, and Ni. 
}
\label{fig:AHC}
\end{figure}

In this subsection, we compute the AHC of $M$Nb$_3$S$_6$ with $M=$Mn, Co, Fe, and Ni.
In most of the states that we consider, there are symmetry reasons that ensure that AHC must be zero. 
In particular, c-AFM, copl-AFM, and ncpl-AFM2 magnetic structures in $M$Nb$_3$S$_6$ remain unchanged under application of the time-reversal symmetry followed by the spatial translation of the lattice. For instance, in the cases of collinear and coplanar antiferromagnetic states, the relevant spatial translation is the one that connects two nearest $M$ sites  with the opposite spin orientations.
On the other hand, noncoplanar AFM states may or may not  break this symmetry, depending on the relative sign of scalar spin chirality in the two magnetic layers~\cite{shindou_orbital_2001, martin_scalar_2008}.

Figure~\ref{fig:AHC} displays the AHC results as a function of the Fermi energy $E_F$ computed for $M$Nb$_3$S$_6$ compounds with the non-coplanar AFM spin texture. For that, we assume rigid bands and vary their occupancy. This allows us to test how sensitive our results are to the precise band alignment, which may not be perfectly captured by DFT, but also possible sensitivity to chemical or electrostatic gate doping.
Consistently with our expectations, out of noncoplanar states ncpl-AFM1 state has finite and large AHC, while ncpl-AFM2 does not.

In the four compounds we studied, the Co and Fe variants show the largest  AHC when the Fermi energy is equal to zero, which is the nominal value for charge-neutral systems. 
Still, the AHC magnitude  depends rather sensitively on the Fermi energy, i.e., the electron filling. We also compute AHC in ncpl-AF1 and ncpl-AF2 states, even though from DFT calculations (Table \ref{tbl:energy2}) we expect the Mn compound to be FM. Also, in the case of Ni the ground state is predicted to be ncpl-AF2, which should have no AHC. The differences of DFT energies are rather small, however, so we cannot rule out that the energy ordering of candidate states can deviate from the DFT predictions.

The sensitivity of the AHC values to the Fermi energy implies the importance of band crossings near the Fermi level. 
In CoNb$_3$S$_6$, the AHC value is $\sim 1.2e^2/h$ without dopings and remains almost flat under electron doping (increasing the Fermi energy).
This can be attributed to the major contribution to the Berry curvature $\Omega(\mathbf{k})$ in CoNb$_3$S$_6$ coming from the vicinity of the high-symmetry  point $K$;  there are no low-energy bands above the Fermi energy at  $K$ point, which makes that contribution insensitive to the Fermi energy.
On the other hand, the hole doping leads to visible reduction of the AHC values.

Typically, in calculations of the Berry curvature $\Omega$, the major contributions to $\Omega$ originate from a small momentum space region.
To identify this region, we compute $\Omega^z(\mathbf{k})$ summed over occupied bands along the high-symmetry directions in Fig.~\ref{fig:non_col_band}.
The sharp peak and valley structures indeed occur in  small regions, but these regions vary depending on the specific transition metal ion $M$.
We also note that  AHC is mostly coming  from the vicinity of $k_z=0$ plane rather than the $k_z=\pi$ plane. 
This can be understood from band structure since the spin-polarized bands are still nearly degenerate along the path in the $k_z=\pi$ plane while the effect of symmetry breaking occurs more prominenty near the $k_z=0$ plane with avoided crossings along the high-symmetry $k-$path.

\begin{figure}[!h]
\includegraphics[width=0.9\linewidth]{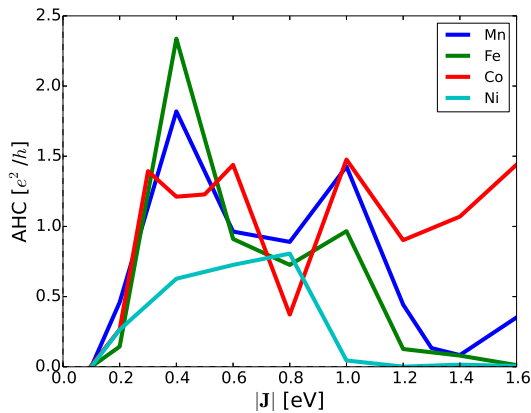}
\caption{Calculated AHC of $M$Nb$_3$S$_6$ as a function of the spin-exchange potential magnitude $|\mathbf{J}|$ for $M$=Mn, Fe, Co, and Ni in the ncpl-AFM1 state.
}

\label{fig:AHC_vs_J}
\end{figure}

Finally, we verify that our AHC results are not very sensitive to the values of $J$ that we used to fit the magnetic band structure.
In Fig.~\ref{fig:AHC_vs_J}, we computed the AHC as a function of ${J}$ in the chiral ncpl-AFM1 state (recall that based on DFT, only Co and Fe compounds are expected to have this ground state).
In all cases, sizable AHC values are obtained when $J$ is in the intermediate range $0.3{\rm \, eV} < J < 0.8{\rm \, eV}$. The zero-temperature values of $J$ for Co and Fe  indeed fall within this range, while it is rather smaller in the case of Ni ($J =0.2$~eV). On the other hand, Mn has very large exchange coupling ($J=1.3$~eV). We recall, however, that the quoted values of $J$ include in themselves the ordered magnetic moment, and therefore the magnitude of $J$ is generally expected to decrease with increasing temperature. This opens an interesting possibility, in the case of the Mn-intercalated compound in particular, that despite the low-temperature state being expected to be c-FM, it may undergo a transition into another, possibly, noncoplanar AFM state with increasing temperature.

\section{Conclusion}
\label{sec:conclusion}

In this work, we  searched among  collinear and non-collinear magnetic configurations for the ground states  of $M$Nb$_3$S$_6$ compounds with $M=$Mn, Fe, Co, and Ni.  
In these materials, the magnetic ions $M$ are far apart and their interactions are predominantly mediated by itinerant electrons. This leads to long-range frustrated exchange interactions, as well as higher order -- multi-spin -- interactions. Even weak higher order interactions generated this way can lift the massive degeneracies common in Heisenberg models, leading sometimes to exotic non-coplanar states instead of the more common coplanar helical states \cite{Solenov2012, Motome2017}.
Indeed, we found that for  FeNb$_3$S$_6$ and CoNb$_3$S$_6$, a non-coplanar AFM structure with uniform scalar spin chirality  has the lowest energy among the plausible candidate states that we considered. In the case of Ni, the lowest energy state is also non-coplanar AFM, but with staggered scalar spin chirality. In contrast, the collinear FM was found to be the ground-state in MnNb$_3$S$_6$. 
We also performed the anomalous Hall conductivity calculations using the band structure obtained from the Wannier Hamiltonian fitted to magnetically ordered states without the spin-orbit coupling.

The obtained results show the AHC on the scale of $\sim e^2/h$ per NbS$_2$ layer, which is comparable with  the experimentally measured AHC values in CoNb$_3$S$_6$. The calculated AHC depends rather sensitively on the chemical potential, indicating that doping may shift the bands  near the Fermi energy and significantly affect the AHC values.
We also studied the effect of a local spin exchange field $\mathbf{J}$ on the AHC calculation.  We found that the intermediate values of $|\mathbf{J}|$  produce the largest AHC. This is indeed the case for both Co and Fe ions.    
The $|\mathbf{J}|$ value obtained from DFT for Ni ion is smaller and thus likely  to produce  weaker AHC even if ordered magnetically  with uniform scalar chirality. 

Given the expected small energy differences between coplanar and non-coplanar states, it is possible that their energies can be relatively rearranged in experiment by, e.g.,  quantum confinement (exfoliation), applying various strain fields, or by doping. Moreover, application of an external magnetic field, via coupling to orbital magnetization \cite{OM2007},  can both align chiral domains in the ncpl-AFM1 state (predicted for $M$ = Co and Fe), and even induce transitions from ncpl-AFM2 (Ni case) to ncpl-AFM1, with a concomitant jump in AHC.

\section*{acknowledgments}
We thank M. Norman for helpful and insightful discussions.
HP, OH, and IM acknowledge funding from the US Department of Energy, Office of Science, Basic Energy Sciences, Division of Materials Sciences and Engineering.
We gratefully acknowledge the computing resources provided on Bebop, high-performance computing clusters operated by the Laboratory Computing Resource Center at Argonne National Laboratory.

\bibliography{main}

\end{document}